\documentclass[traditabstract]{aa} 
\usepackage{txfonts,graphicx,url,natbib}

\begin{document}

\title{Constraining the shaping mechanism of the Red Rectangle through spectro-polarimetry of its central star}

\author{M. J. Mart\'{\i}nez Gonz\'alez\inst{1,2}, A. Asensio Ramos\inst{1,2}, R. Manso Sainz\inst{1,2}, R. L. M. Corradi\inst{1,2} \and F. Leone\inst{3}}

\institute{Instituto de Astrof\'\i sica de Canarias, 38205, La Laguna, Tenerife, Spain; \email{marian@iac.es}
              \and
           Departamento de Astrof\'{\i}sica, Universidad de La Laguna, E-38205 La Laguna, Tenerife, Spain
              \and
           Dipartimento di Fisica e Astronomia, Universit\'a di Catania, Sezione Astrofisica, Via S. Sofia 78, I-9512, Catania, Italy}
             
\date{Received ---; accepted ---} 
 
\abstract{We carried out high-sensitivity spectropolarimetric observations of the central star of the Red Rectangle 
proto-planetary nebula with the aim of constraining the mechanism that gives its biconical shape. 
The stellar light of the central binary system is linearly polarised since it is 
scattered on the dust particles of the nebula. 
Surprisingly, the linear polarisation in the continuum is aligned with one of the spikes of the biconical 
outflow. Also, the observed Balmer lines as well as the Ca\,{\sc II} K lines are polarised. These 
observational constraints are used to confirm or reject current theoretical models for the shaping mechanism 
of the Red Rectangle. We propose that the observed polarisation is very unlikely generated by a uniform biconical stellar wind. 
Also, the hypothesis of a precessing jet does not completely match the observations since it will require a jet aperture 
larger than that of the nebula.}

\authorrunning{Mart\'{\i}nez Gonz\'alez et al.}
\titlerunning{Constraining the shaping mechanism of the Red Rectangle}
\maketitle

\section{Introduction}

The Red Rectangle is a proto-planetary nebula with a biconical shape extending away 
from the central object HD 44179. The central object is known to be a single
line spectroscopic binary system composed of a secondary
star that accretes material from the primary post-AGB star
forming an accretion disc. The nature of the secondary star
is still a matter of debate: \cite{menshchikov_02} propose that
it is a white dwarf although \cite{waelkens_96} and \cite{witt_09}
consider that it is a low-mass main sequence star. The accretion
disc of the secondary star is supposed to accelerate a secondary
wind along its polar axis forming a jet. 

The central stellar system is obscured by an optically 
thick torus and is observed from the scattering of the light in dust particles. This 
scattering produces two bright hyperbolic arcs with the southernmost one brighter than the 
northernmost one. The reason why one lobe is brighter that the other is still not known 
\citep{cohen_04} but it is possibly related to the scattering physics and/or the 
distribution of dust particles in both hemispheres (note that 
the inclined torus obscures a larger portion of the northern hemisphere and the scattered light comes 
from larger distances from the central system). Expanding discs around post-AGB stars are rather common 
but the Red Rectangle is the only object with a rotating disc \citep{bujarrabal_05} plus the 
expected expansion \citep{bujarrabal_castro_13}.

Two main physical mechanisms have been proposed to explain
the shape of most bipolar nebulae. The most popular
one states that in a close binary system angular momentum
provides a natural preferential axis \citep{morris_81, morris_87}. A
second mechanism involves the presence of a mostly dipolar
magnetic field where the matter can be collimated at the poles
of the field. However, at present, a clear detection of such
fields has not been found \citep{leone_11,jordan_12}.

In order to shape the Red Rectangle, two main scenarios
have been proposed: one that assumes that the secondary
wind is a well collimated jet inclined by 30$^\circ$ 
(with respect to the symmetry axis of the nebula) that precesses
each 17.6 years \citep{soker_05,velazquez_11}, and another
one that assumes a bipolar outflow (or wide jet) with the same
opening as the nebula \citep{morris_81, morris_87, soker_00, 
thomas_13, icke_81, koning_11}, these three last ones 
considering the special case of a biconical
outflow. Spectro-polarimetry is a valuable tool that allows
us to differentiate between the two scenarios. The physical
mechanism that shapes this nebula must explain the observed
light polarisation. Here we present linear spectro-polarimetry of the 
central stellar system of the Red Rectangle. We compare these observations 
with the theoretical expected linear polarisation from different mechanisms 
proposed to give the biconical shape to this object.

\section{Observations and data reduction}

The data consist of full-vector spectro-polarimetry of the central binary system of the Red Rectangle (HD 44179) 
at blue wavelengths, between 365 and 510 nm, approximately. The observations were taken on 
September 4th 2012 using the FORS2 polarimeter \citep{fors2} attached to the Very Large Telescope (VLT) 
at Cerro Paranal. 
We used the 1200B grism and a slit width of 0.5$''$ to achieve a spectral sampling of 
0.036 nm at 450 nm. The total integration time for each Stokes parameter was 8 min, which allowed us 
to obtain a polarimetric signal-to-noise ratio of about 1250. The slit was oriented 55$^\circ$ 
from the celestial North to the East, close to the parallactic angle at that moment. The inclination 
of the slit does not introduce any bias in the observed polarisation since the central object \hbox{HD 44179} 
is unresolved. As seen in Fig. 7 in \cite{cohen_04}, the maximum of the scattered light 
of the central object is enclosed in about $\sim 0.2"$.

In this Letter, we focus on the study of the linear polarisation that carries information about the 
unresolved geometry of the central system. The circular polarisation pattern, which allows the 
detection of magnetic fields if present, has been analyzed by our group \citep{andres_14, leone_14} showing 
that there is no clear evidence of an organized magnetic 
field in the central binary system of the Red Rectangle. 

The FORS2 polarimeter is composed of two retarders (a quarter wave plate and a half wave plate), 
and a beam splitter (the analyzer). 
For convenience, we rotated the retarders to obtain the fractional polarisation 
(the polarisation relative to the intensity) according to the following modulation scheme in
one of the beams produced by the analyzer: 
$F_I+F_j$, $F_I-F_j$, $F_I-F_j$, $F_I+F_j$, with $j={Q,U}$. Note that 
the orthogonal state of polarisation for each modultation step is recorded simultaneously in
the other beam position. With the available measurements, we compute the 
fractional polarisation $F_j/F_I$ as \citep{bagnulo_09}:
\begin{equation}
\frac{F_j}{F_I}=\frac{1}{2}\sum_{i=1}^2\frac{I^1_i/I^2_i}{I^1_{i+1}/I^2_{i+1}},
\label{eq:demodulation}
\end{equation}
where $I^1$ and $I^2$ denote the intensity measured in each one of the beams, while the index $i$ represents 
each one of the modulation steps. This modulation scheme allows us to efficiently correct, to 
first order, the flatfield, differential 
defocusing, and different seeing conditions during the different exposures. An additional advantage
is that it allows us to compute the null spectra \citep{donati_97}, which gives an idea of the residual errors in the data:
\begin{equation}
\frac{N_j}{F_I}=\frac{1}{2}\sum_{i=1}^2(-1)^i\frac{I^1_i/I^2_i}{I^1_{i+1}/I^2_{i+1}}.
\end{equation}
Given that the spectrograph is not perfectly stabilized, we need to re-align the spectra in
wavelength for each modulation state. We do this by using a calibration arc lamp exposure. To avoid
possible spurious signals \citep{bagnulo_09}, we use only a single wavelength calibration for all the modulation
states.
Additionally, given that the spectra are not aligned with the borders of the
detector, we use the flatfield image to compute the position of
the beams at each wavelength. Once they are aligned, we integrate each beam along
the slit direction and then use Eq. (\ref{eq:demodulation}) to obtain the
fractional polarisation. The reason for doing the demodulation in this order is that the
error propagation is simpler and more advantageous. 

In general, it is quite difficult to know a-priori which of the two beams on the detector
contains $F_I + F_j$ or $F_I - F_j$. The reason is that this requires us to know exactly
the position of the axis of the analyzer, which is not the case in many instruments. 
Note that, if one switches both beams, the
resulting polarisation will be rotated by 90$^\circ$. To calibrate the two
beams, we observe
a linear polarisation standard star whose polarisation degree
and angle are known. This observation can also help to estimate
if there is any induced instrumental polarisation. In our observing
run we observed the Hiltner 652 star, which has a degree
of polarisation of 5.72$\pm$0.02\% and a polarisation angle of 179.8$\pm$0.1$^\circ$
as observed with a filter in the Johnson B band\footnote{\texttt{http://www.eso.org/sci/facilities/paranal/\\
instruments/fors/inst/pola.html}}. 
Given that these are broad band calibrations, we compute from our data the 
degree and angle of polarisation from the integrated Stokes $F_Q$ and $F_U$ wigthed by the Johnson B 
filter transmission, $B(\lambda)$, as:
\begin{equation}
\left \langle \frac{F_i}{F_I} \right \rangle = \frac{\int_{\lambda_1}^{\lambda_2}d\lambda\, B(\lambda)\, 
\frac{F_i}{F_I}}{\int_{\lambda_1}^{\lambda_2} d\lambda\, B(\lambda)},
\end{equation}
where the polarisation degree and angle are defined as
\begin{eqnarray}
\frac{F_P}{F_I}&=&\sqrt{\left(\frac{F_Q}{F_I}\right)^2 + \left(\frac{F_U}{F_I}\right)^2}\\ \nonumber
\theta&=&\frac{1}{2} \arctan{\frac{F_U}{F_Q}}.
\end{eqnarray}
We obtain $\langle F_P/F_I \rangle=5.9\pm0.6$\% and 
$\theta=0.0\pm0.5^\circ$, which are compatible with the tabulated values inside
the error bars. This means that the instrumental polarisation is negligible and that the
instrument rotation is properly taken into account in the data reduction.

\begin{figure*}
\centering
\includegraphics[width=0.8\textwidth]{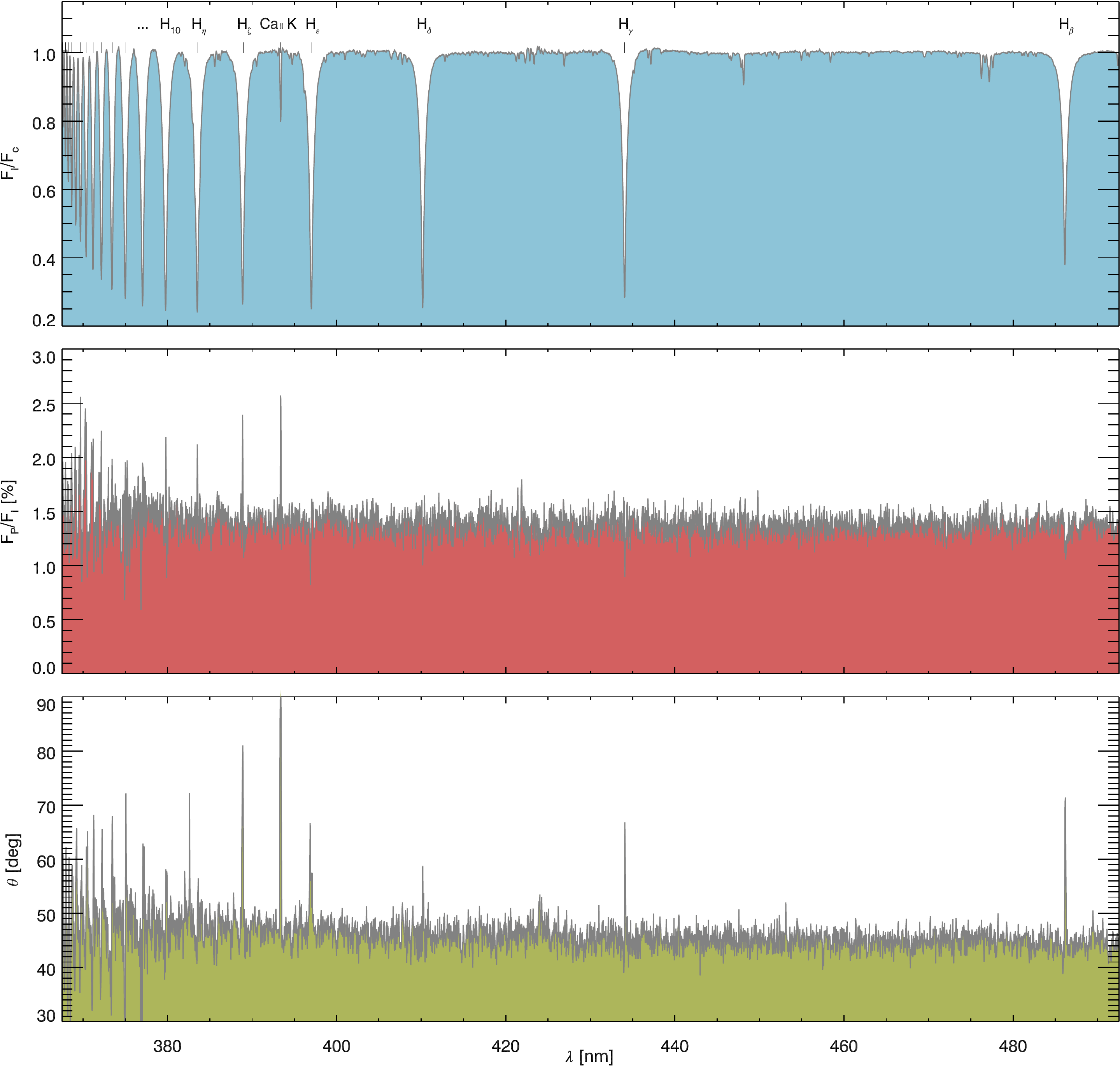}
\caption{From top to bottom, the blue spectra of the intensity flux, 
the polarisation degree and the polarisation angle of the central star of the red rectangle.
Note that the mode of the $F_P/F_I$ distribution is the 
standard deviation of Stokes $F_Q$ and $F_U$ parameters, which means that if both $F_Q$ and $F_U$ 
are zero, the $F_P/F_I$ will have a value of 0.08\%.}
\label{fig:full_spectrum}
\end{figure*}

Given that we are dealing with absolute polarimetry, it is important to 
check the influence of interstellar polarisation along the line-of-sight. 
\cite{reese_96} estimate (using field
stars) that the degree of polarisation of the interstellar component
is at most 0.2\%. Since they observe a degree of polarisation of 2.2\% in the continuum,
this means that only $\sim$9\% of their observed continuum polarisation in the Red Rectangle can be
attributed to interstellar polarisation.
Unfortunately, they do not compute the estimated rotation angle produced
by the interstellar medium, so we try to provide an upper limit to this effect.
The influence of the interstellar polarisation on the polarisation of the object is
given by
\begin{eqnarray}
\left(\frac{F_Q}{F_I}\right)_\mathrm{obs}&\sim& \left(\frac{F_Q}{F_I}\right)_\mathrm{obj} + \left(\frac{F_P}{F_I}\right)_\mathrm{IM} \cos{2\theta_\mathrm{IM}}\\
\left(\frac{F_U}{F_I}\right)_\mathrm{obs}&\sim& \left(\frac{F_U}{F_I}\right)_\mathrm{obj} + \left(\frac{F_P}{F_I}\right)_\mathrm{IM} \sin{2\theta_\mathrm{IM}},
\end{eqnarray}
where the subindex ``obs'' represents the observed Stokes parameters,
i.e., the addition of the polarisation coming from
the object (labeled with ``obj``) and the polarisation coming from the interstellar
medium (labeled with ``IM'').
This formula is a good approximation if the polarisation of the interstellar medium is small. Also, if this happens, 
we can substitute the object polarisation (that is unknown)
with the observed one. Then, to first order, we have:
\begin{eqnarray}
\left(\frac{F_Q}{F_I}\right)_\mathrm{obs} &\sim& \left(\frac{F_Q}{F_I}\right)_\mathrm{obs} + \left(\frac{F_P}{F_I}\right)_\mathrm{IM} \cos{2\theta_\mathrm{IM}}\\
\left(\frac{F_U}{F_I}\right)_\mathrm{obs}&\sim& \left(\frac{F_U}{F_I}\right)_\mathrm{obs} + \left(\frac{F_P}{F_I}\right)_\mathrm{IM} \sin{2\theta_\mathrm{IM}},
\end{eqnarray}
 Assuming the value of 
\cite{reese_96} for the polarisation degree of 0.2\% and using all possible values of $\theta_\mathrm{IM}$,
we find that the maximum influence on the polarisation angle inferred from $F_Q/F_I$ and $F_U/F_I$ is only $\pm4^\circ$.

Our data offers another possibility to compute an upper limit to the polarisation of the
interstellar medium. The Ca\,{\sc II} H line at 3968.5 \AA, a transition between
$J_{\mathrm{up}}=J_{\mathrm{low}}=1/2$ levels, cannot be polarised by scattering
processes\footnote{Under anisotropic pumping, the populations of the $m_J>0$ and $m_J<0$ 
magnetic sublevels are not the same. The Ca\,{\sc ii} H line has $J_\mathrm{up} = J_\mathrm{low} = 1/2$ and 
almost 100\% of the Ca\,{\sc ii} has no hyperfine structure. 
Then, it can never have population imbalances between its magnetic sublevels and then produce 
linear polarisation in the emission process.}.
Therefore, all the photons emitted in the Ca\,{\sc ii} H line are not polarised,
so that the degree of polarisation will tend to zero. Consequently, if
the degree of polarisation of the Red Rectangle at the Ca\,{\sc II} H 
line is non-zero, then the observed absolute polarisation is coming from
the interstellar medium. However, in general, the emission
of spectral lines is not enough to fully depolarise (except for
some very strong lines) and the polarisation we observe in the Ca\,{\sc II} H 
line (0.8\%) is an upper limit for the polarisation of the interstellar
medium. The same applies to the angle of polarisation in this
line. Using the previous conditions, a very conservative maximum error in the polarisation
angle is $\sim\pm17^\circ$. We conclude
that the measured angle of polarisation of the Red Rectangle
is well defined. 

\begin{figure*}
\center
\includegraphics[width=0.55\textwidth]{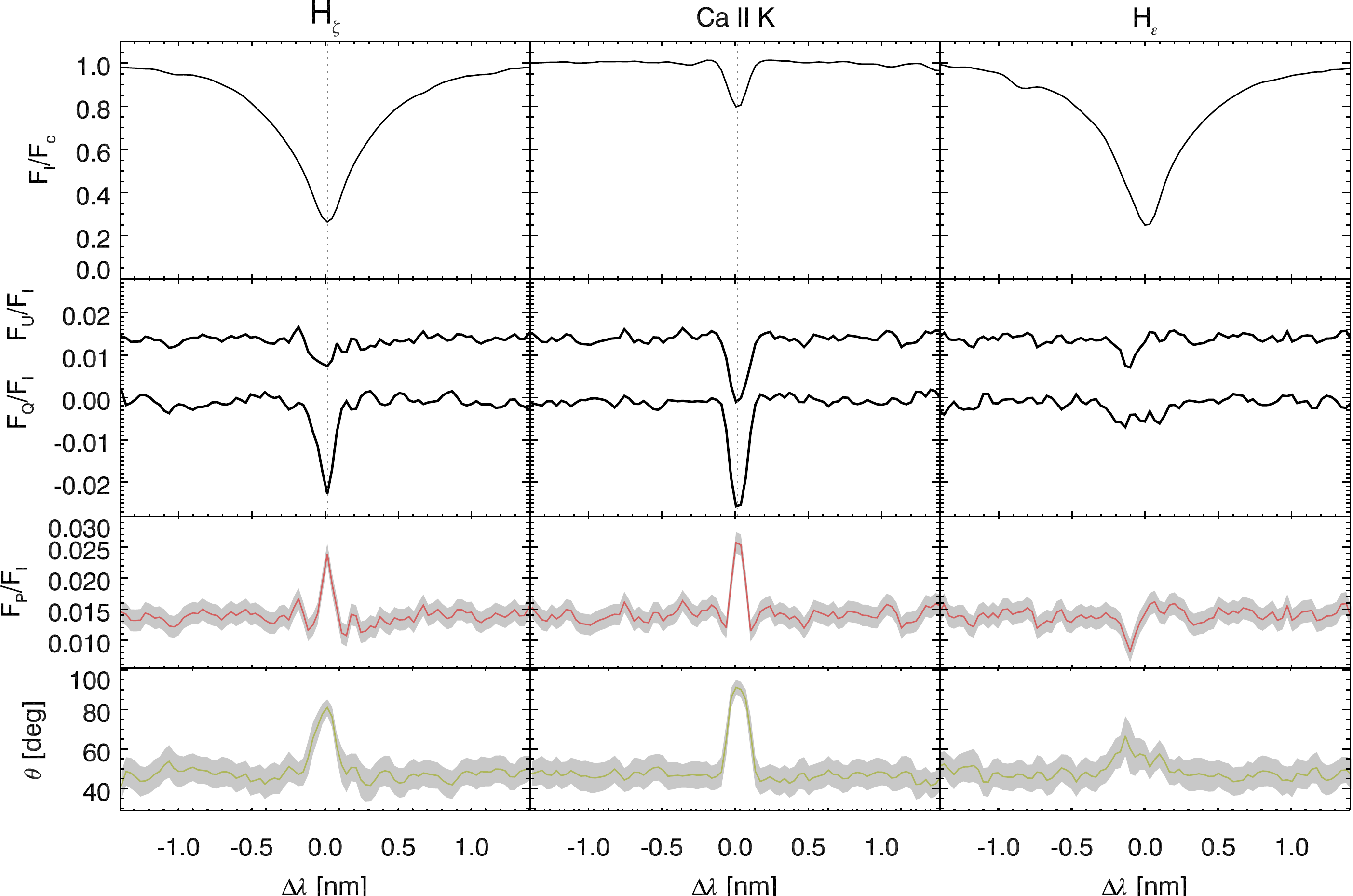}\vspace{0.3cm}
\includegraphics[width=0.55\textwidth]{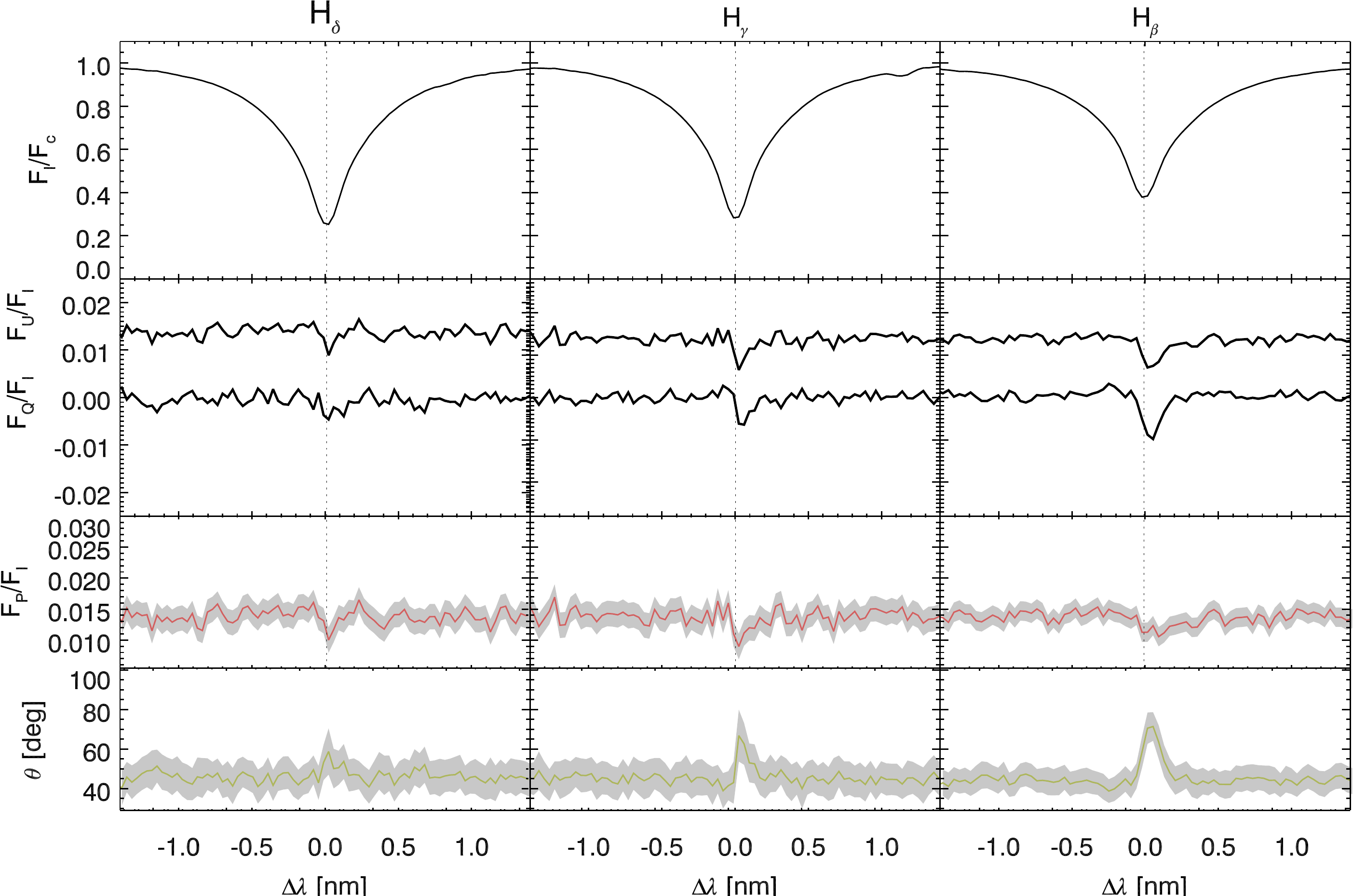}\vspace{0.3cm}
\includegraphics[width=0.55\textwidth]{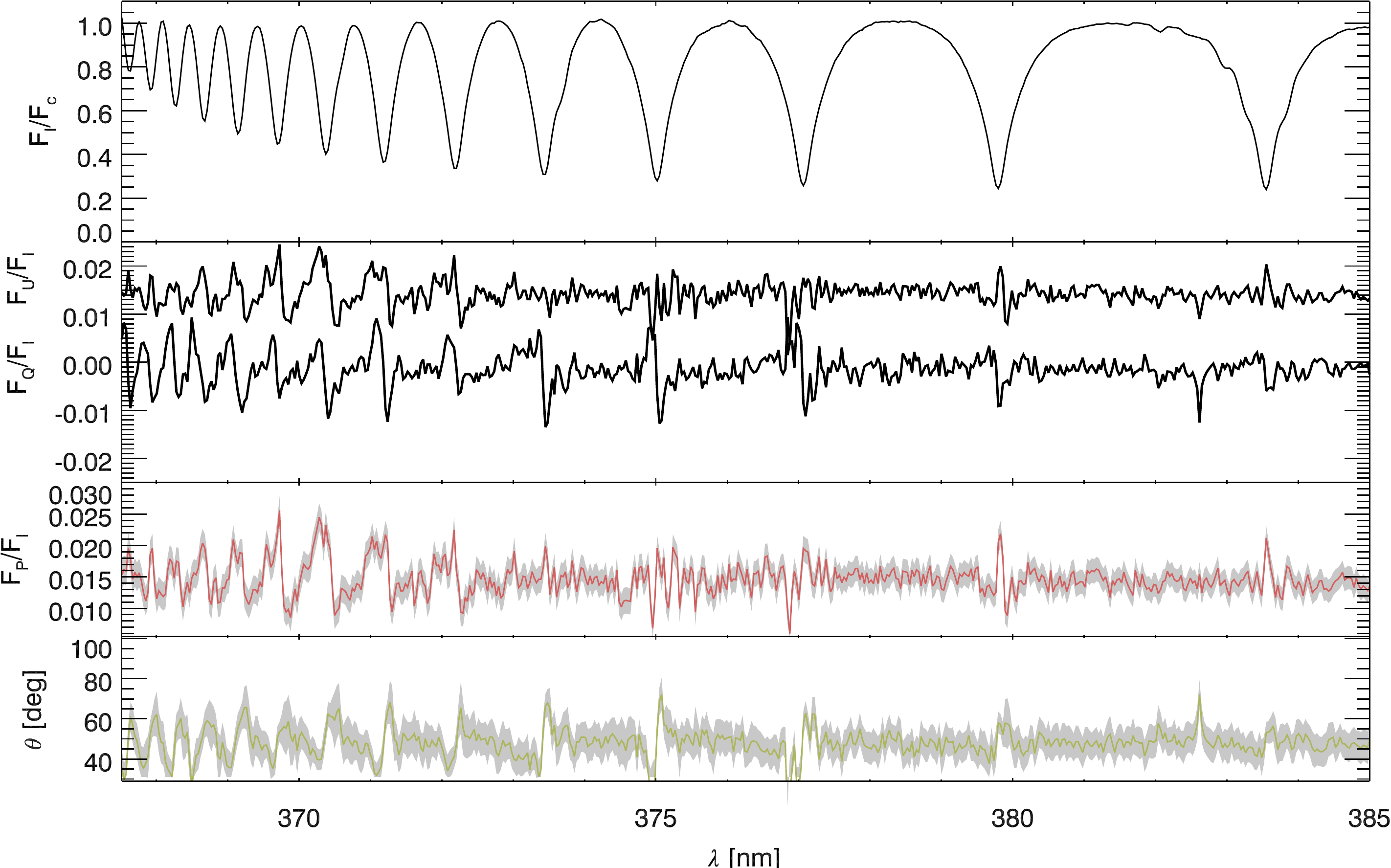}
\caption{Close look to the Balmer lines and the Ca\,{\sc II} K line. 
From top to bottom, the spectra of the intensity flux, the flux of the Stokes $Q$ and $U$ 
parameters in terms of the intensity flux, the polarisation degree, and the polarisation angle. 
The shadowed areas represent the 2-sigma confidence level for these derived quantities.
\label{fig:zoom_lines}}
\end{figure*}

\section{Results}


Figure \ref{fig:full_spectrum} displays the observed intensity flux spectrum (upper panel), 
the polarisation degree (middle panel) and angle (lower panel). Contrary to the polarisation 
degree, that does not depend of the chosen reference system used to measure
the Stokes parameters, the polarisation angle, 
does depend on the reference system. We define $\theta = 0^\circ$ at the
celestial N-S direction \citep[for an easier comparison with][]{reese_96},
increasing towards the East. Note that the polarisation angle
does not depend on the intensity flux.

\begin{figure*}[!t]
\center
\includegraphics[width=0.45\textwidth]{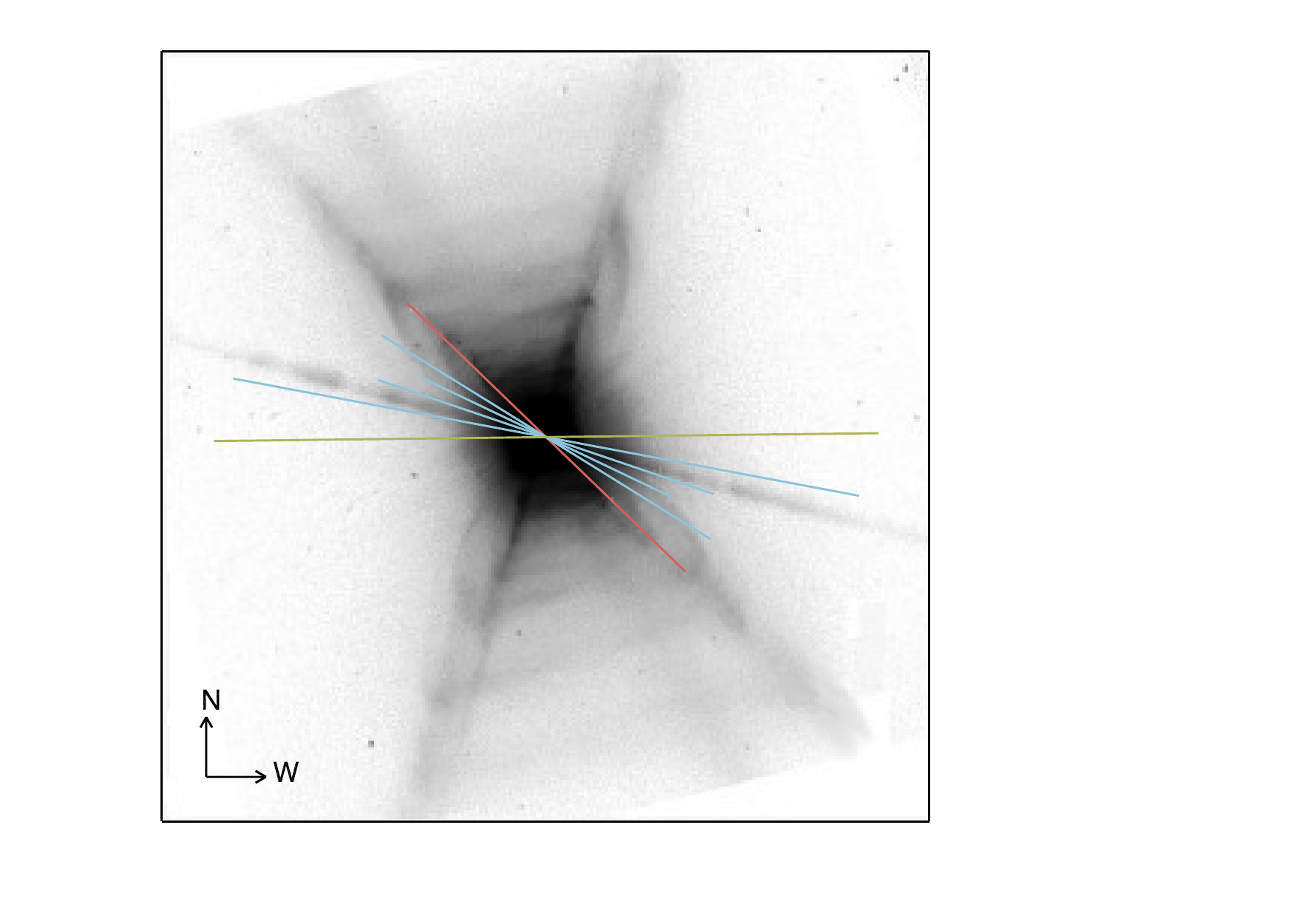}
\includegraphics[width=0.4555\textwidth]{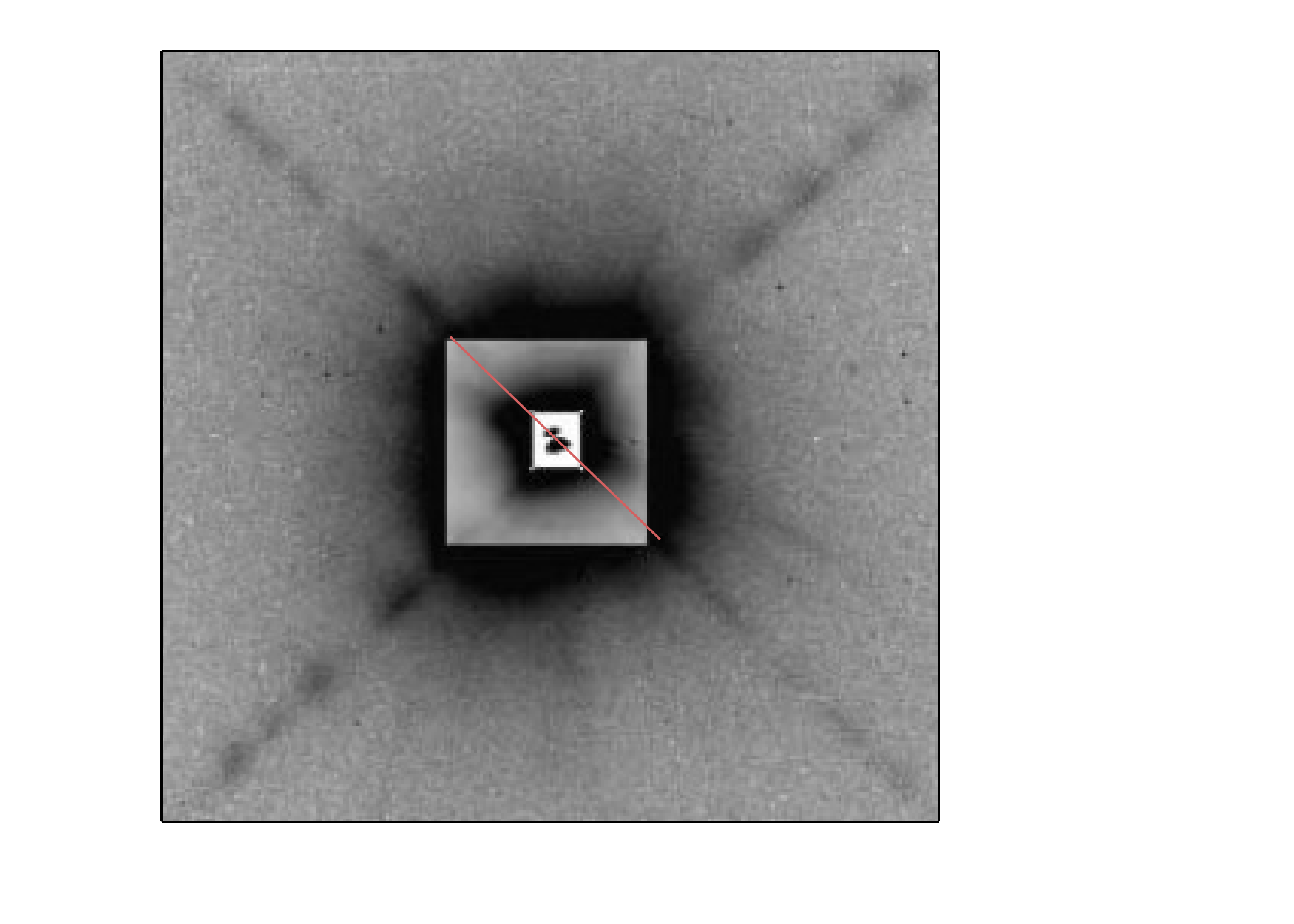}
\caption{The backgroud image represents the Red Rectangle protoplanetary nebula as 
seen in the red \citep[left panel; taken from][]{koning_11} and in the
blue \citep[right panel; taken from][]{cohen_04}. The lines represent the direction 
of the polarisation of the continuum (red), the Balmer lines (blue) and the
Ca\,{\sc II} K line (green). The size of the lines represents the degree polarisation as respect 
to the continuum value.\label{fig:angles}}
\end{figure*}

The continuum
is polarised at $\sim$1.4\% with an angle of $\sim45^\circ$ with respect to
the celestial N. Given that the symmetry axis of the conical structure is
inclined 10$^\circ$ with respect to the celestial N, the polarisation
makes an angle of 35$^\circ$ with respect to the symmetry axis of the nebula, 
coinciding, surprisingly, with one of the spikes of the
nebula. 
\cite{reese_96}, 17.8 years before than our observation, measured a continuum degree of
polarisation of 2.2\%. The measured polarisation angle was $\sim$43$^\circ$, very
similar to our result. It is likely that the
50\% difference between the values of the degree of polarisation
at the continuum is due to the different sensitivity of the
two data sets, with our data being much more sensitive. However, 
we need more observations to check if this variation is real 
or not.

Observations of continuum linear polarisation of the outer parts of the Red Rectangle 
have been previously reported by \cite{perkins_81} in the 
red (6100 - 7500 \AA) and blue wavebands (4000 - 5000 \AA). In both cases, they 
observe the polarisation angle following a centrosymmetric pattern, as expected 
from the scattering of the light from the central object in the walls of the bicones. The 
same result is found by \cite{gledhill_09} in the 
near infrared. The compatibility of these two works and our result is 
a matter of spatial scales. If the physical mechanism that breaks the symmetry plays a major role 
in the central parts of the nebula (has a small spatial extent of about 1"), the polarisation we measure 
is an effect of this symmetry breaking. If we move to the outer 
nebula, the linear polarisation is blind to this smaller scale mechanism and becomes 
a reflect of the symmetry of the large scale nebula. However, note that in Figure 1 of Gledhill et al. (2009), 
the polarisation in the nbL filter (the one that better "resolves" the central parts of the nebula) 
shows polarisation angles that are not 
centrosymmetric (at about $\sim 60^\circ$) at 1" surrounding the central object.

Figure \ref{fig:zoom_lines} shows a closer look at the intensity flux, Stokes
$Q$ and $U$ parameters, and total polarisation and angle for the 
Balmer lines and the Ca\,{\sc II} K line. The shadowed areas in polarisation
and in the polarisation angle represent the $2\sigma$ uncertainty levels produced
by photon noise using the formula from \cite{bagnulo_09}. 
Most of the Balmer lines, notably H$_\beta$, H$_\gamma$,
H$_\epsilon$ , H$_\zeta$, and H$_\eta$ have linear polarisations clearly rotated with respect
to the continuum by 20$^\circ-45^\circ$. It is interesting to point out that the scattering process for
the lines H$_\epsilon$, H$_\delta$, H$_\gamma$ and H$_\beta$ has only modified the angle and
not the polarisation degree. The H$_\epsilon$ line is blended with the Ca\,{\sc II} 
H line wich can not be polarised, hence, it can be affected by
an artificial reduction of the polarisation angle. The H$_\delta$ line,
however, is just slightly rotated with respect to the continuum
($\sim 5^\circ-10^\circ$), in contrast to the rest of Balmer lines. The Ca\, {\sc II} 
K line presents the largest degree of polarisation and also the
largest angle ($\sim 85^\circ$). Given that the transmission of the isntrumental system below 380 nm is
poor, we prefer not to conclude anything until it is observed with better signal to noise ratio, even
though some features are seen. Unfortunately, our data coverage does not reach 
H$_\alpha$, but we point out that \cite{reese_96} observe an excess of polarisation in 
this line of 7.9$\pm$0.6\% with an polarisation angle of 102$\pm$2$^\circ$ with respect to the celestial North.

Figure \ref{fig:angles} displays the directions of the polarisation of the
continuum (red), the Balmer lines (blue) and the Ca\,{\sc II} K
line (green) superimposed to the image of the Red Rectangle
proto-PN as observed in the red wavelengths (left panel)
and in the blue (right panel). The nebula looks more spherical
in the blue wavelengths and has a smaller size \citep{perkins_81, cohen_04}. 
As stated above, the continuum polarisation is aligned
with one of the spikes of the proto-PN. The linear polarisation of the spectral lines are
rotated with respect to this direction by $20^\circ-45^\circ$ 
to the East direction, H$_\zeta$ presenting a linear polarisation that is 
almost perpendicular to the spike of the proto-PN that is not aligned with the
linear polarisation of the continuum.

\section{Discussion}

The fact that we detect non-zero continuum polarisation, and a rotation of the 
angle with respect to the equatorial plane, 
can only be explained if there is a certain degree of symmetry
breaking in the system. Several options are available and we discuss
them in the light of the observations.
The most obvious one is that the symmetry breaking is related to 
the binarity. Using the information provided in \cite{thomas_13} 
for the ephemerids of the binary system, the observations of
\cite{reese_96} was carried out at phase $\phi=0.36$, while ours was done at
phase $\phi=0.84$, almost in opposite orbit phases. 
The propagation of the error on the period estimation of the binary system
might make the two phases compatible. However, recently \cite{witt_09} confirmed a period of $318 \pm 1$
day, i.e., an error of only 0.3\%. Therefore, we have to conclude that the orbital phases
of our observations and those from \cite{reese_96} are completely 
different. Considering Rayleigh scattering, we expect a 1\% modulation of the 
degree of polarisation between the periastron and the apastron. Unfortunately, this is 
still not detectable considering our observational capabilities. 

\begin{figure}[!t]
\center
\includegraphics[width=\columnwidth]{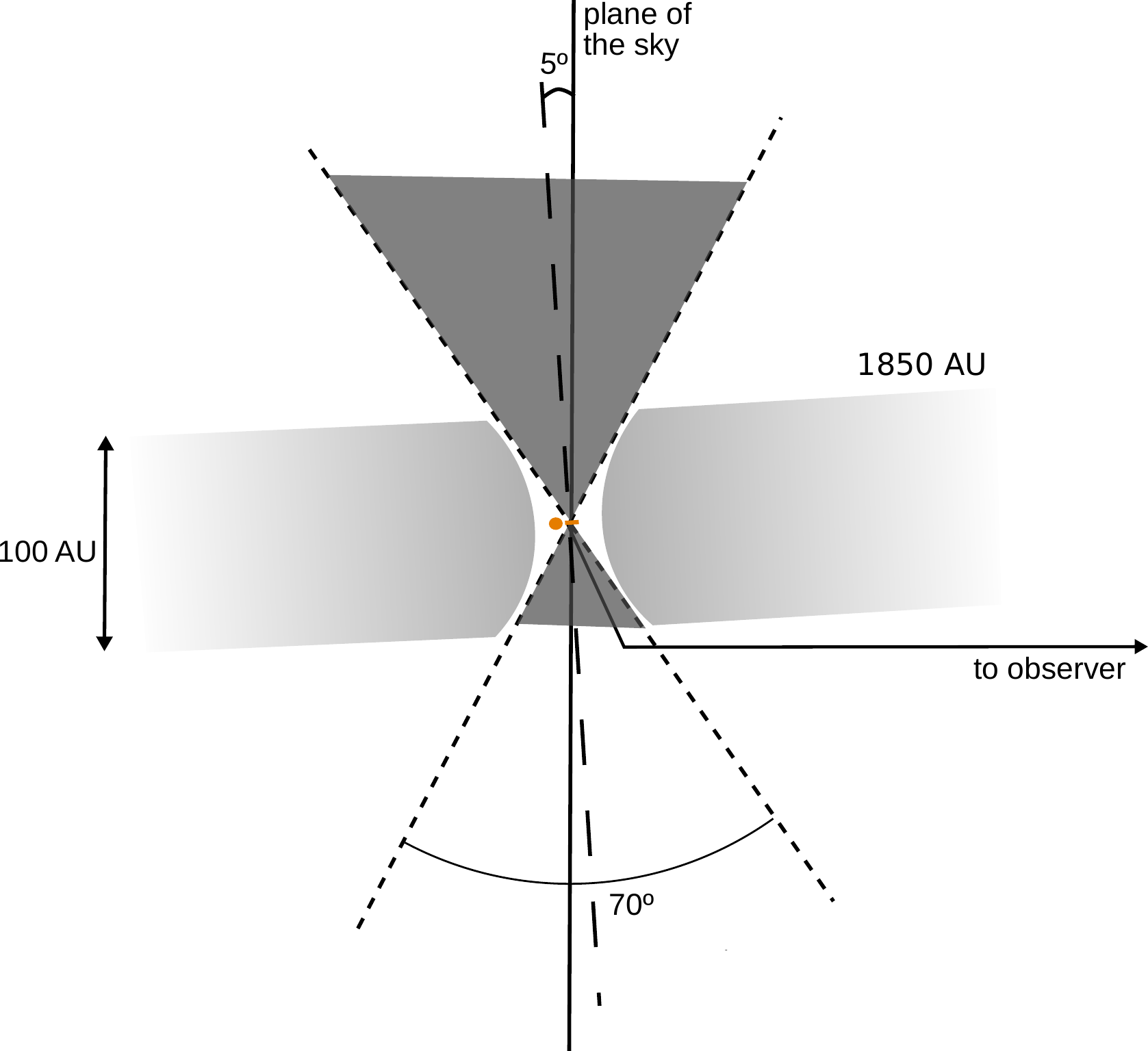}
\caption{Schematic drawing of the unresolved center of the Red Rectangle, 
which is inclined by 5$^\circ$ with respect to the line-of-sight (LOS).
The plane of the sky is perpendicular to the sheet, and the line-of-sight is represented in red. 
The spikes that are observed in the red are represented as dashed black lines. The
central binary system is represented by an orange circle (the AGB) and an orange rectangle (the accretion disc). 
The light grey area around the central system is
the optically thick torus, which has an outer radius estimation of 1850 AU, an inner radius 
(the cavity of the binary system) of 14 AU, and a thickness of about
100 AU \citep{bujarrabal_13}. The dark grey area represents the area of the Red Rectangle which is hidden to the observer 
by the circunbinary disc. Note that the disc is so large that
about 161 AU height as measured from the plane of the binary system is obscured.}
\label{fig:geometry}
\end{figure}

The presence of a thick dusty torus \citep{bujarrabal_13} and the fact
that it is slightly tilted (see Fig.\ref{fig:geometry}) is another source of symmetry breaking. The dusty
disc allows us to view the light of the binary star only by scattering processes. As can be seen 
in Fig. \ref{fig:geometry}, the radius of the disc occults regions closer than $\sim$211 AU 
to the binary star in the upper cone, and regions closer than $\sim$50 AU in the lower cone. 
Since both the density of scatterers and the radiation field decrease from the central binary
system, we can safely consider that the scattering is fundamentally coming from the lower half of 
the circunbinary
disc, i. e., at a distance of $\sim$50 AU as measured from the equatorial plane of the binary system.
In absence of any other mechanism, and for symmetry reasons,
this would generate linear polarisation that should be parallel or perpendicular to
the symmetry axis of the nebula. In order to explain 
the rotation of $\sim$45$^\circ$ with
respect to the celestial N, we need to invoke other symmetry breaking mechanisms.

\begin{figure}[!t]
\center
\includegraphics[width=0.5\textwidth]{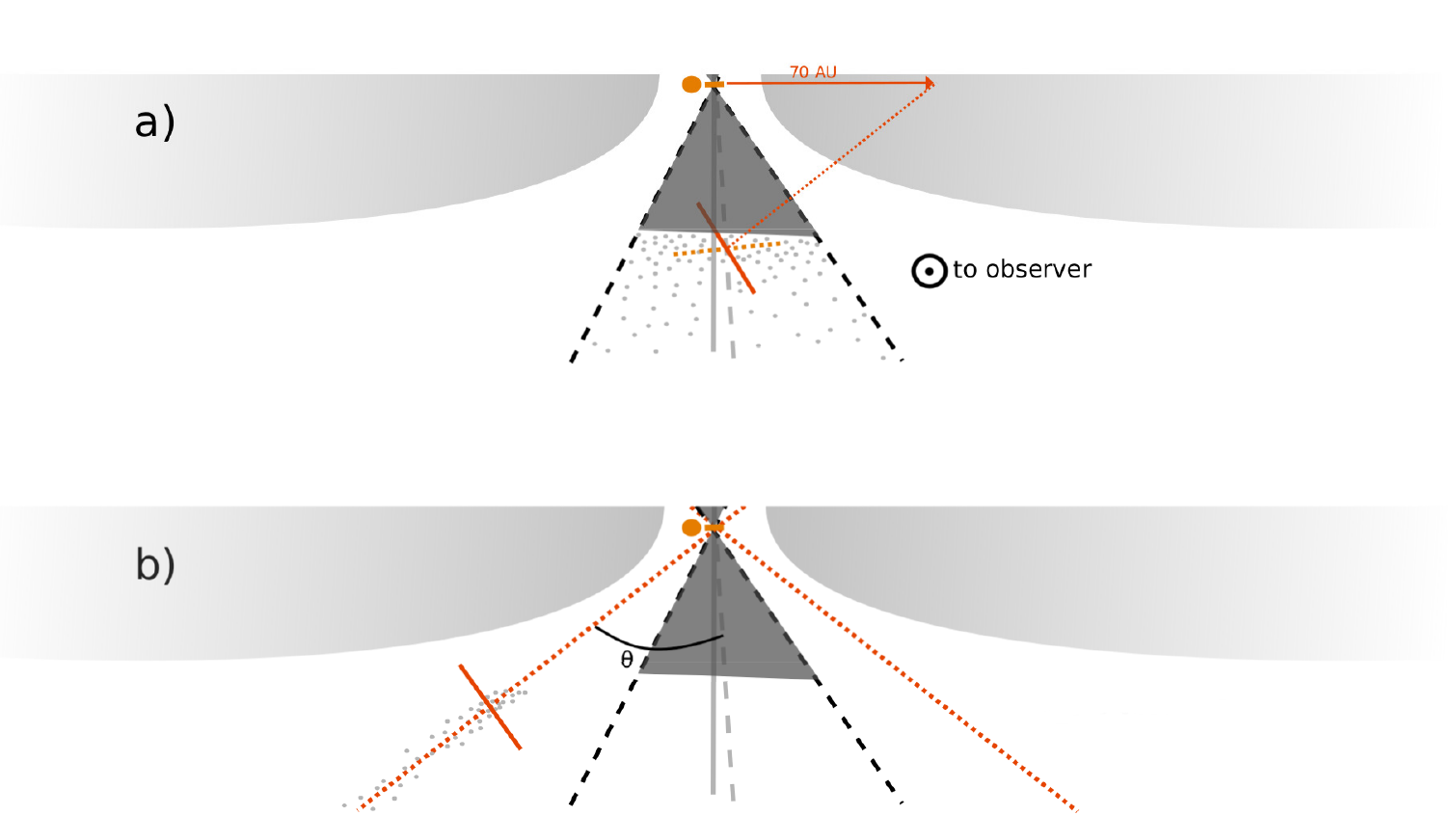}
\caption{Possible scattering mechanisms producing the polarisation in the continuum that we
observe. We only display one cone (the lower cone but mirrored with respect to the
orbital plane) since the scattered light is mostly generated there (i.e. the larger the 
distance from the central object, the lower the density and the intensity of the radiation 
field). Grey dots represent the scatterers. The spikes of the nebula are traced in black dashed lines, the central
system is represented in orange, the horizontal line representing the accretion disc. 
The observed linear polarisation is displayed in red. 
Case a) depicts the scattering produced at all points of the cone (i.e.
the wide jet mechanism). The expected theoretical polarisation in this case is represented as 
a short-dashed orange line. 
Case b) displays the case when the scattering occurs at the aperture $\theta$ that 
gives a polarisation compatible with the observed one.\label{fig:mechanisms}}
\end{figure}

In the light of the previous considerations, we can think of two mechanisms that are possible to explain the 
observations. The two possibilities
are depicted in Fig. \ref{fig:mechanisms}.
The first one is that the scattering takes place in the bipolar flow (case a, see 
upper panel of Fig. \ref{fig:mechanisms}). The
second one is that it happens in a dense collimated jet (case b). 
In case a), if the binary system is located
at the center of the nebula, we expect the polarisation of the
continuum to be in the equatorial plane of the torus-binary
system \citep[e.g.][]{daniel_82, cohen_82, harries_96}. 
In order to produce the observed rotation which produces polarisation
parallel to the spike of the nebula, the binary system would have to be
decentered by \hbox{$d=70$ AU} (taking into account the geometric considerations shown in Fig. \ref{fig:geometry}). 
Considering that the
star separation is always smaller than 1 AU \cite{thomas_13} and that the inner radius of the torus
is 14 AU \citep{menshchikov_02}, we find that a bipolar flow alone can not explain the
observed continuum polarisation in the Red Rectangle. 
Additionally, if a bipolar flow is taking place, we expect the polarisation to be
modulated by the orbital period of the binary system. Yet,
both \cite{reese_96} and our observations, presenting very different
phases, show the same angle of polarisation.

In case b), where we assume a dense collimated jet, the scattering
will take place mainly at the jet. Consequently, the polarisation
will be perpendicular to the scattering plane formed by the jet
and the line-of-sight. According to this picture (see lower panel of Fig. \ref{fig:mechanisms}) and
our measurements, we can estimate that the opening angle of this jet should be $\sim$55$^\circ$. 
If the jet is precessing, we would get a biconical structure with a full aperture of 110$^\circ$.
The full aperture of the biconical walls of the Red Rectangle measured from imaging data vary from 
40$^\circ$ close to the central source to 80$^\circ$ at larger distances \citep{cohen_04}. On average, 
the opening angle we infer is 20$^\circ$ larger that the semi-aperture of the nebula
(measured from the symmetry axis to the nebula) and 25$^\circ$ larger than the expected aperture of a 
precessing jet computed by \cite{velazquez_11}. In spite of this difference, we note that the Red Rectangle 
has a larger aperture (about 90$^\circ$) as observed in polarised light at optical wavelengths 
\citep[see Fig. 2 in][]{perkins_81}. Also, it is important to note that our considerarions are purely
geometric in nature and that any radiative transfer effect on the spectral lines can
easily compensate for these differences in the polarisation angle. 
A precessing jet puts constraints on the modulation of the observed polarisation
with the rotational period. Unfortunately, there are
17.8 years since the observations of \cite{reese_96}, a similar period 
to the one needed in \cite{velazquez_11} to shape the
Red Rectangle with a precessing jet (17.6 years). In order to confirm or
discard the hypothesis of a precessing jet, we plan to spectro-polarimetrically
survey the Red Rectangle in the forthcoming
years.

A feature that we find intriguing is the fact that the linear spectro-polarimetry of 
the Red Rectangle reveals that the hydrogen Balmer 
lines and the doubly ionized Ca K line are polarised with respect to the continuum, and 
are additionally rotated with respect to the continuum by different amounts depending 
on the spectral line. Only resonant scattering, i.e., 
polarisation in spectral lines, can produce a rotation of the polarisation angle. 
Also, the fact that the Ca\,{\sc ii} K is the most polarised line and is, in turn, 
the most shallow line in the intensity spectrum, reinforces the assumption that the observed 
signals are produced by scattering in spectral lines. Scattering polarisation in spectral 
lines is induced by the presence
of population imbalances and quantum coherences (atomic level polarisation) between
magnetic sublevels caused by the absorption of anisotropic
radiation \citep[e.g.,][]{egidio}, which is mainly controlled by the amount of
anisotropy of the radiation field. Note that any other mechanism 
(two different offset sources for the continuum and the spectral lines, or stray light) 
can not explain the observed polarisation angles and would hardly polarise spectral lines 
(in general, spectral lines would appear as a depolarisation of the continuum).

The rotation of resonance scattering polarisation with respect to the symmetry axis of the nebula 
indicates that there is another preferential axis that breaks the apparent symmetry of the nebula. 
Several mechanisms can produce such a rotation. A magnetic field naturally imposes another 
preferential axis, modifying the polarisation signals 
leading to a rotation of the plane of polarisation through the Hanle effect. To get such a rotation, 
the magnetic field strength should be in the so-called Hanle critical field, 
which is $\sim 1-100$ G for the hydrogen lines \citep[e.g.,][]{jiri_11} and $\sim 1-120$ G for 
the Ca\,{\sc ii} K line. Under the assumption of a dipolar magnetic field in the central region 
of the Red Rectangle, we expect the magnetic field strength to decrease as the cube of the distance. 
Therefore, the Hanle effect has to be discarded as a source of the rotation of the linear polarisation 
because, in order to have Hanle effect at $\sim$50 AU, the magnetic field strength at 1 AU would
have to be $0.1-10$ MG. 

Another way of producing a rotation of the polarisation is by optical depth
effects, which has been proposed in the past to explain polarimetric observations
of SiO masers \citep{asensio_masers05}. The specific orientation of the polarisation depends on
the balance between radiation escaping preferentially along the flow (giving polarisation
perendicular to the symmetry axis of the nebula) or preferentially along the orbital plane
of binary star (giving polarisation parallel to the symmetry axis of the nebula). Local
density accumulations produce a modification of the anisotropy of the radiation field that
can lead to local rotations of the polarisation of the emitted radiation. Additionally,
a large velocity gradient along a certain direction produces an enhancement of the anisotropy
along this direction via the Doppler brightening effect \citep{carlin12}. This could be
related to the inherent mass-loss of the star.

The observed angle of polarisation at the blue continuum reveals 
that a wide jet is unlikely
shaping the Red Rectangle \citep[c.f.][]{morris_81, morris_87, soker_00, 
thomas_13, icke_81, koning_11}. A large mass inhomogeineity
\citep[like a collimated jet][]{soker_05, velazquez_11} could explain the shape of the nebula
if it is orientated along a position angle of about 55$^\circ$. With gathering more data, our future efforts 
will be put to understand the physical mechanisms 
that generate the polarisation properties of this 
proto-planetary nebula.

\begin{acknowledgements}
We are very grateful to S. Bagnulo for helpful advise on the data reduction, and to 
an anonymous referee whose comments have strengthen the conclusions of this work. 
Support by the Spanish Ministry of Economy and Competitiveness 
through projects AYA2010--18029 (Solar Magnetism and Astrophysical Spectro-polarimetry) and Consolider-Ingenio 2010 CSD2009-00038 
are gratefully acknowledged. AAR also acknowledges financial support through the Ram\'on y Cajal fellowships.
\end{acknowledgements}


\end{document}